# Complexity of Two-Dimensional Patterns

Kristian Lindgren [*]   Cristopher Moore [†]   Mats Nordahl [‡]

July 23, 2018

**Abstract**

In dynamical systems such as cellular automata and iterated maps, it is often useful to look at a *language* or set of symbol sequences produced by the system. There are well-established classification schemes, such as the Chomsky hierarchy, with which we can measure the complexity of these sets of sequences, and thus the complexity of the systems which produce them.

In this paper, we look at the first few levels of a hierarchy of complexity for two-or-more-dimensional patterns. We show that several definitions of "regular language" or "local rule" that are equivalent in $d = 1$ lead to distinct classes in $d \geq 2$. We explore the closure properties and computational complexity of these classes, including undecidability and **L**, **NL** and **NP**-completeness results.

We apply these classes to cellular automata, in particular to their sets of fixed and periodic points, finite-time images, and limit sets. We show that it is undecidable whether a CA in $d \geq 2$ has a periodic point of a given period, and that certain "local lattice languages" are not finite-time images or limit sets of any CA. We also show that the entropy of a $d$-dimensional CA's finite-time image cannot decrease faster than $t^{-d}$ unless it maps every initial condition to a single homogeneous state.

[*]Institute of Physical Resource Theory, Chalmers University of Technology, S-412 96 Göteborg, Sweden frtkl@fy.chalmers.se
[†]Santa Fe Institute, 1399 Hyde Park Road, Santa Fe, New Mexico 87501 moore@santafe.edu
[‡]Institute of Theoretical Physics, Chalmers University of Technology, S-412 96 Göteborg, Sweden tfemn@fy.chalmers.se, and the Santa Fe Institute mgn@santafe.edu

# 1 Introduction

## 1.1 One-dimensional languages in physics

Consider a dynamical system, for instance an iterated map $F$ acting on some space $U$. If we partition $U$ into $k$ subsets $U_1, U_2, \ldots U_k$, then for any initial point $x$ we can write down a sequence $(a_t)$ of symbols describing which subset it falls into at each time-step; i.e. $a_t = j$ if $F^t(x)$ is in $U_j$. This sequence then describes a coarse history or *itinerary* of $x$. If the map is invertible, $a$ can also be extended backwards, producing a bi-infinite sequence.

A logical question to ask, then, is: what possible sequences can the system produce? In other words, for what sequences is there an $x$ with that sequence as its itinerary? This set of sequences is called the *symbolic dynamics* of the map $F$, and can be a very useful way to classify the system; often the partition can be chosen so that the map between points and sequences is one-to-one, allowing us to enumerate its periodic points [20] and calculate quantities like entropies, escape rates and Liapunov exponents [1].

As another example, consider a cellular automaton (CA) in one dimension. This is a dynamical system on sequences where each site is updated according to some local rule, as a function of its state and those of its neighbors; for instance, suppose the state at each site is 0 or 1, and $F(a)_i = f(a_{i-1}, a_i, a_{i+1})$ for some Boolean function $f$. Then we can ask a variety of questions, such as: what sequences $(a_i)$ are in the image of $F$ after one iteration? After two? What sequences are fixed points, i.e. $F(a) = a$? What sequences are periodic points, in that $F^t(a) = a$ for some $t$? What sequences map onto the zero state, $F(a) = (0)$? And what points are in the limit set, the intersection of the images of $F^t$ for all $t > 0$?

All these questions refer, as does the symbolic dynamics question above, to sets of sequences, or *languages*. Clearly some languages are more complex than others; the set of sequences $\{10^p1 \,|\, p \text{ prime}\}$ of two 1's separated by a prime number of 0's, for instance, is clearly more complex than the set of words containing an equal number of 0's and 1's, which in turn is more complex than the set of sequences where two 1's never occur consecutively. This qualitative notion of complexity is formalized by the *Chomsky hierarchy* (e.g. [23]), in which languages are classified by the different types of machines needed to recognize or generate them. Originally proposed by Noam Chomsky as a set of models of natural language, this hierarchy has since been taken up by computer scientists and others seeking to quantify the notion of complexity.

The basic Chomsky classes are called, from simplest to most complex, *regular*, *context-free*, *context-sensitive*, and *unrestricted*; these correspond to increasingly powerful kinds of machines, at the top of which sits the Turing machine (e.g. [41]) which, according to the Church-Turing thesis, is computationally universal.

In fact, examples all up and down this hierarchy can be found in dynamical systems theory. The languages generated by many simple hyperbolic systems are regular; this corresponds to the existence of a finite Markov partition for their dynamics [20]. At phase transitions such as the period-doubling fixed point, however, they can have a complicated scale-invariant structure, and belong to an intermediate class called *indexed context-free* [9]; and the iteration of smooth maps in the plane can correspond to universal Turing machines [43].

Similarly, the image of a cellular automaton after a finite number of timesteps is regular [60], as is the set of fixed points; but limit sets can be context-free, context-sensitive, or the complement of the halting set of a Turing machine [24].

The purpose of this paper is to introduce the reader to an analogous hierarchy of two-dimensional "languages" or patterns of symbols. This hierarchy turns out to be much richer than in one dimension, in that several equivalent definitions of regular languages generalize in subtle ways to become distinct classes in $d \geq 2$. We hope that such a hierarchy will allow us to more clearly discuss issues of complexity in spin systems, cellular automata, coupled map lattices, and other systems in two or more dimensions.

To provide background, we review the definitions of regular and context-free languages in $d = 1$. Readers interested more in this subject should consult [23] or another text on the theory of languages and automata.

## 1.2 Equivalent descriptions of regular languages

The recognition machine is a paradigmatic object in language theory. It is fed a word as input, and accepts or rejects it according to whether or not that word is in the language. The simplest kind of machine is the *deterministic finite-state automaton* (DFA): it consists of a box with an internal state in some finite set $S$, which reads a tape on which a candidate word is written.

Suppose the language is written in some set of symbols or *alphabet* $A$. Then the DFA reads the tape from left to right, letting its state at each step depend its old state and the symbol it is currently reading according to a *transition function* $F : A \times S \to S$. After it reaches the end of the tape, it accepts the word if its final state is in some subset $S_{\text{accept}}$ of $S$, and rejects it otherwise.

For example, consider the language $L_{\overline{bb}}$ consisting of words of the alphabet $\{a, b\}$, where the only rule is that no two $b$'s may occur consecutively. This language is accepted by a DFA with three internal states, $A$, $B$, and $R$ for 'Reject', with $F$ described by

|   | $a$ | $b$ |
|---|---|---|
| $A$ | $A$ | $B$ |
| $B$ | $A$ | $R$ |
| $R$ | $R$ | $R$ |

Then we start in state $A$, accept if we end up in $A$ or $B$, and reject if we end up in $R$.

We define a language as *regular* if it is recognized by some DFA. This embodies the idea of a language where only a finite amount of memory is required to recognize it.

There are several ways to generalize DFA's in an effort to make them more powerful: for instance, we can consider *non-deterministic* finite-state automata or NFA's. Their dynamics consists of a function $F : A \times S \to \wp(S)$ whose values are subsets of $S$, giving the machine one or more choices at each step of which state to adopt. We say that an NFA accepts a word if there exists some set of choices which leads it to an accepting state.

Non-deterministic machines are in general more powerful than deterministic ones, since their definition of acceptance allows them to test many possibilities simultaneously: for instance, it is believed that many problems can be solved in polynomial time non-deterministically but not deterministically (i.e, $\mathbf{P} \neq \mathbf{NP}$).

However, in the case of finite-state automata, the NFA is no more powerful than the DFA. Create a DFA whose states are subsets of $S$, $S' = \wp(S)$. Then let $F'(a, s') = \cup_{s \in s'} F(a, s)$, and define a state $s' \in S'$ as accepting if it contains some accepting state, i.e. $S'_{\text{accept}} = \{s' \in \wp(S) \,|\, s' \cap S_{\text{accept}} \neq \emptyset\}$. Clearly this DFA will accept the word if and only if the NFA has an accepting trajectory. So NFA's can be simulated by DFA's, and can recognize the same class of languages (although the equivalent DFA might be exponentially larger).

Another seemingly more powerful machine we could consider is a *two-way* finite-state automaton (2DFA or 2NFA) which can move both left and right on its input tape — surely an advantage, since it can go back and recall previous characters of the input. But in fact this is no more powerful than the one-way kind: construct an automaton with states representing a record of all the times and states in which the 2-way FA visited a given place on the tape. These *crossing sequences* are finite, since a 2FA with $n$ internal states can visit each site no more than $n$ times without falling into a loop. A local matching rule, enforceable by an NFA, then ensures that crossing sequences at adjacent sites are consistent; details are given in [23].

So in one dimension we can say that

$$\text{DFA} = \text{NFA} = \text{2DFA} = \text{2NFA}$$

since all these machines recognize regular languages.

The class of regular languages is also preserved under a variety of operations. Let a *homomorphism h* map the alphabet $A$ onto some smaller alphabet $h(A)$, collapsing some symbols and losing information in the process; mapping both $a$ and $b$ onto $a$, for instance. If we do this to a regular language, the resulting language is again regular, since an NFA can guess which of the original symbols in $A$ it should use.

Regular languages have a number of other descriptions, including:

**1.) Transition graphs.** If we label nodes with states and edges with tape symbols, DFA's and NFA's can be written as directed graphs. Such a graph has a transition matrix $M_a$ for each symbol $a$, and the number $N(l)$ of allowed words of length $l$ has the leading behavior $\lambda^l$ where $\lambda$ is the largest eigenvalue of their sum $M = \sum_a M_a$.

**2.) Regular expressions.** If we read '+' as 'or', multiplication as concatenation, $w^*$ as '0 or more repetitions of $w$', and $\epsilon$ as the empty string, regular languages are those which can be expressed with a finite formula using these operators. For instance, $L_{\overline{bb}}$ can be written $(a+ba)^*(\epsilon+b)$ or $(\epsilon+b)(aa^*b)^*a^*$.

**3.) Regular grammars.** A grammar [23] consists of a set of symbols $V$, including a start symbol $S$ and a set $P$ of production rules $\alpha \to \beta$, where $\alpha$ and $\beta$ are strings of symbols. The language generated by the grammar consists of the strings that can be derived from $S$ by applying the production rules in arbitrary order.

Regular grammars only have productions of the form $A \to wB$ and $A \to w$, where $A$ and $B$ are symbols and $w$ is a string of *terminal* symbols that cannot change or create more symbols. Thus the variables move to the right, leaving a string of terminals behind them. For instance, the grammar

$$\begin{aligned} S &\to aS \text{ or } bA \text{ or } \epsilon \\ A &\to aS \text{ or } \epsilon \end{aligned}$$

generates the strings of $L_{\overline{bb}}$.

**4.) Positive pumping lemmas.** A useful property of regular languages is the *pumping lemma*, which states that any sufficiently long string $x$ in a regular language $L$ can be written as $x = yzw$ where $yz^n w \in L$ for all $n \geq 0$. This can often be used to show that a language is non-regular. There are positive versions of the pumping lemma that are both necessary and sufficient for a language to be regular [37].

**5.) Rational formal power series.** Consider the formal sum

$$\epsilon + a + b + aa + ab + ba + aaa + aab + aba + baa + bab + aaaa + \ldots$$

of all the words in $L_{\overline{bb}}$. This sum can be viewed as the expansion of the rational function

$$\frac{1}{1 - a - ba}(1 + b)$$

where $a$ and $b$ are non-commuting variables and $1 = \epsilon$. The power series of rational functions in non-commuting variables correspond exactly to the regular languages; there is a beautiful theory of such series outlined in [52].

**6.) Equivalence classes.** In a language $L$, we can define two words as equivalent, $u \sim v$, if they can be followed by the same set of suffixes, i.e. $ux \in L$ if and only if $vx \in L$. Then a language is regular if and only if $\sim$ has only a finite number of equivalence classes, which correspond to the states of the smallest DFA that recognizes it. Thus the smallest DFA is unique up to isomorphism.

## 1.3 Finite complement languages

The language $L_{\overline{bb}}$ is an example of a *finite complement* (f.c.) language, in that it can be defined by a finite list of forbidden substrings, namely $\{bb\}$. Thus the language can be described in a purely local way; equivalently, we could list the allowed blocks $\{aa, ab, ba\}$ of length 2. Sets of infinite sequences defined in this way are called *subshifts of finite type* (e.g. [32]).

Clearly any finite complement language is regular, but the reverse is not the case: for instance, $(a^*ba^*c)^*$ is regular but not f.c., since an infinite set of substrings $ba^*b$ and $ca^*c$ would have to be excluded. Whether the last non-$a$ was a $b$ or a $c$ is a 'hidden state', obscured by arbitrarily large blocks of $a$'s.

However, every regular language is a homomorphism of some f.c. language. If we label the edges of the transition graph with distinct symbols, we get $(a^*bd^*c)^*$, which is f.c.; its allowed blocks are $aa$, $ab$, $bd$, $dd$, $dc$ and $ca$. We can ensure that we start and end with the proper states by allowing only $\sharp a$ and $c\sharp$, where $\sharp$ is a putative blank character which lies outside the word. By mapping $d$ to $a$, we recover the original set. (In symbolic dynamics, homomorphisms of subshifts of finite type are called *sofic systems* [58].)

In any case, homomorphisms of f.c. languages, which we will call h(LLL)'s below, are yet another way to define regular languages in one dimension.

## 1.4 Context-free languages

In several places, we will use *context-free* languages, the second lowest level in the Chomsky hierarchy [23]; they properly contain the regular languages. A language is context-free if it is recognized by a *push-down automaton* (PDA), a finite-state machine with access to a stack memory. On reading an input symbol, it can read (and pop) the top symbol of the stack, update its internal state, and/or push new symbols onto the stack. It accepts if it starts and ends with an empty stack.

The canonical context-free language is the Dyck language $\{\epsilon, (), (()), ()(), (())(), \ldots\}$ of well-formed words of parentheses; another example is the set $\{a^n b^n\}$ of words consisting of a block of $a$'s followed by an equal number of $b$'s. Both of these languages are context-free but not regular.

## 2 Two-dimensional languages

How do various definitions of regular language generalize in two or more dimensions? We will show that DFA's, NFA's and homomorphisms of finite complement languages, which were all equivalent to regular languages in one dimension, become distinct classes of increasing subtlety, Even finite complement languages (which we call Local Lattice Languages, or LLL's) are capable of structure much more subtle than in the one-dimensional case.

In essence, each of these classes represents a different concept of *locality* in a system's structure. Apparently the distinction between local and global in two or more dimensions is actually quite tricky, and different attempts to capture it lead to very different sets of languages.

## 2.1 Notation

If $A$ is a finite alphabet, let $\Sigma = A^{\mathbb{Z} \times \mathbb{Z}}$ be the set of infinite two-dimensional *pictures* or arrays of symbols in $A$, and let $\Sigma_{m,n} = A^{m \times n}$ be the set of $m \times n$ blocks. In analogy with one-dimensional languages, we will usually construct languages of finite blocks, $L \subset \cup_{m,n} \Sigma_{m,n}$; however, we are also interested in sets of infinite configurations, $L_\infty \subset \Sigma$. If these are closed and translationally invariant, they are called *subshifts* as in the one-dimensional case (e.g. [32]).

To translate back and forth between finite and infinite blocks, we introduce extension and restriction operators $E$ and $R$. If $L$ and $L_\infty$ are languages of finite and infinite configurations respectively, let

$$E_\infty(L) = \{B \in \Sigma \,|\, \forall b \subset B : b \in L\}$$

$$R(L_\infty) = \{b \in \cup_{m,n} \Sigma_{m,n} \,|\, \exists B \in L_\infty : b \subset B\}$$

where by $b \subset B$ we mean that $b$ is a sub-block of $B$. Then $E_\infty(L)$ is the set of infinite configurations containing only blocks in $L$, and $R(L_\infty)$ is the set of finite blocks appearing in infinite configuations in $L_\infty$. To extend finite blocks to larger but still finite blocks, we define

$$E(L) = \{B \in \cup_{m,n} \Sigma_{m,n} \,|\, \forall b \subset B : b \in L\}$$

which is the set of finite blocks containing blocks in $L$.

Note that $E$ and $R$ are by no means inverses of each other! The set of *infinitely extensible* blocks $R(E_\infty(L))$ is typically a proper subset of $L$, and is often of a higher order of complexity than $L$. In fact, even its non-emptiness is undecidable, as we will see below.

We will often be interested in the number of allowed blocks of a certain size; we will call the number of $m \times n$ rectangles in $L$ the *growth function* of $L$, $N(m,n)$. The *entropy per site* is then

$$\sigma = \lim_{m \to \infty, n \to \infty} \frac{\log N(m,n)}{mn}$$

if this limit exists. (Since we are simply counting states without a notion of measure, this is often called the *topological* entropy.)

Finally, we need to say something about boundary conditions. We imagine our finite blocks as surrounded by a special blank symbol $\sharp$. By interacting with these $\sharp$'s, our various recognition schemes can detect the block's edges and corners.

## 2.2 Local Lattice Languages, or LLL's

Suppose the only allowed $2 \times 2$ blocks in a 2-d language of $\heartsuit$'s and $\spadesuit$'s are

and their rotations. Alternately, we could say that the block 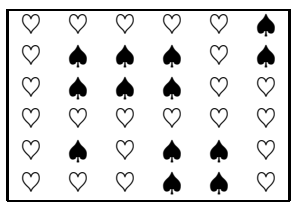 and its rotations are excluded. Then the only allowed configurations consist of rectangular blocks of $\spadesuit$'s floating in a sea of $\heartsuit$'s, such as

We call this language $L_{\text{rect}}$, and discuss it further in Section 3. In general, we say

**Definition.** A two-dimensional language $L$ is a *local lattice language* (LLL) if there exists a finite set of blocks $L_{\text{allowed}}$ such that $L = E(L_{\text{allowed}})$.

In other words, $L$ can be defined by listing a finite number of blocks of a given size or shape, and demanding that pictures in $L$ contain those blocks and no others. The diameter of the largest block in $L_{\text{allowed}}$ is called the *range* of $L$. Clearly, LLL's are analogous to the finite complement languages defined above. We can give several physical examples:

**1.) Defect-free ground states of lattice Hamiltonians with local interactions.** If configurations exist where the local Hamiltonian is minimized everywhere, then the neighborhood(s) which minimize it form the allowed blocks of an LLL. Conversely, every LLL is the ground state of some local lattice Hamiltonian, which assigns a zero energy to allowed blocks and a positive energy to others.

Even if the ground states are frustrated, in that they do not locally minimize the Hamiltonian, they can often (but not always [38]) be represented by an LLL of larger range. For instance, the set of ground states of the antiferromagnet on the triangular lattice is an LLL where the allowed triangles have two $\uparrow$'s and one $\downarrow$, or vice versa; this defines a 3-point Hamiltonian with the same ground states which is locally minimizable.

**2.) Space-time histories of 1-d cellular automata or Turing machines.** If $f$ is a one-dimensional nearest-neighbor CA rule $a'_i = f(a_{i-1}, a_i, a_{i+1})$, then if we allow blocks of the form [a|b|c / f(a,b,c)] the LLL with simulate the CA's evolution from row to row.

In particular, the CA rule can simulate a Turing machine, where special states correspond to the machine's head and internal states, while others correspond to its tape symbols [35]. Then its input appears along the top row, and we can use the LLL to require that it halts or doesn't halt before it reaches the bottom. Thus simple questions about 2-d LLL's can be equivalent to the Halting Problem; this will be our main source of undecidability.

3.) **Fixed and periodic points in 2-d cellular automata.** We will show this in Section 3.

Just as many statistical mechanics models can be exactly solved in one dimension but not in two, 2-d LLL's are often much more subtle than their one-dimensional counterparts. Consider for example the LLL where $\boxed{1\ \ 1}$ and $\boxed{\genfrac{}{}{0pt}{}{1}{1}}$ are excluded; a lattice gas where no two adjacent sites may be occupied. In one dimension this is just $L_{\overline{bb}}$ again, and the entropy per site is $\sigma = \log \tau$ where $\tau = (\sqrt{5}+1)/2$ is the golden mean. However, although $\sigma$ has been calculated to high accuracy in the two-dimensional case [40, 11], it is not known exactly; one the hexagonal lattice, on the other hand, an analytic solution exists [4].

As an example of a zero-entropy LLL, consider the rule where every $2 \times 2$ block must contain an even number of up spins — for instance, the ground state of a 4-point interaction, $H = -K \sum_\Box a_1 a_2 a_3 a_4$ where the $a_i$ are $\pm 1$ and $K > 0$. Allowed configurations consist of the product of horizontal and vertical stripes of ↑'s and ↓'s of arbitrary width, such as

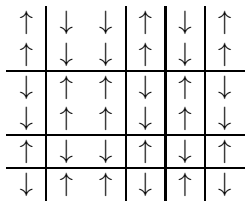

Such a configuration is determined by its topmost row and leftmost column, so the number of allowed $m \times n$ blocks is $N(m,n) = 2^{m+n-1}$ and the entropy per site is zero.

An LLL can enforce local topological properties. For instance, let our alphabet be $\{0, \nwarrow, \nearrow, \searrow, \swarrow\}$. If we exclude blocks where a path branches or ends, i.e. rotations, reflections and reversals of $\boxed{\nearrow\ \nwarrow}$, $\boxed{\genfrac{}{}{0pt}{}{\ \ \swarrow}{\nearrow\ }}$, and $\boxed{\begin{smallmatrix}0 & 0\\ \nearrow & 0\end{smallmatrix}}$, blocks will contain only closed loops or paths that end at the boundary.

Note that we also allow $L_{\text{allowed}}$ to have ♯'s in it, in order to detect the picture's edges; for instance, we can prevent paths from beginning or ending at the boundary by forbidding rotations and reflections of $\boxed{\sharp\ \ \nearrow}$ and $\boxed{\sharp\ \ \nwarrow}$. This forces all the paths to be closed in the block's interior.

We can also get scale-invariant behavior in two dimensions, which local rules could never provide in one — for instance, suppose that our alphabet is $\{0, 1\}$, and that of every block of 3 sites of the shape $\boxed{\genfrac{}{}{0pt}{}{x}{x\ x}}$, either 1 or 3 of the $x$'s must be 0's so that they sum to zero mod 2. If we also forbid $\boxed{\begin{smallmatrix} & \sharp \\ \sharp & 0 \end{smallmatrix}}$ so that the upper-left corner is a 1, we get the mod-2 Pascal's Triangle:

| 1 | 1 | 1 | 1 | 1 | 1 | 1 | 1 |
|---|---|---|---|---|---|---|---|
| 1 | 0 | 1 | 0 | 1 | 0 | 1 | 0 |
| 1 | 1 | 0 | 0 | 1 | 1 | 0 | 0 |
| 1 | 0 | 0 | 0 | 1 | 0 | 0 | 0 |
| 1 | 1 | 1 | 1 | 0 | 0 | 0 | 0 |
| 1 | 0 | 1 | 0 | 0 | 0 | 0 | 0 |
| 1 | 1 | 0 | 0 | 0 | 0 | 0 | 0 |
| 1 | 0 | 0 | 0 | 0 | 0 | 0 | 0 |

If we create some choice by allowing the non-0's to be 1 or 2, $N(n,n) = 2^{n^{\log 3/\log 2}}$ whenever $n$ is a power of 2. Again the entropy per site is zero. In general, it is clear that the growth function of an LLL can have a wider variety of functional forms than in one dimension.

One last amusing example — let the allowed blocks be

$$\left\{ \begin{array}{|cc|} \hline x & 0 \\ x & 0 \\ \hline \end{array}, \begin{array}{|cc|} \hline x & 1 \\ x & 1 \\ \hline \end{array}, \begin{array}{|cc|} \hline x & 0 \\ x & 1 \\ \hline \end{array}, \begin{array}{|cc|} \hline x & 1 \\ \overline{x} & 0 \\ \hline \end{array} \right\}$$

where $x = 0$ or 1, and $\overline{x} = \text{not}(x)$. Then these blocks generate a counting machine, where the last one represents carrying a digit. If we further require

$$\left\{ \begin{array}{|c|} \hline \sharp \\ 0 \\ \hline \end{array}, \begin{array}{|c|} \hline 1 \\ \sharp \\ \hline \end{array}, \begin{array}{|cc|} \hline \sharp & x \\ \hline \end{array}, \begin{array}{|cc|} \hline x & \sharp \\ \overline{x} & \sharp \\ \hline \end{array} \right\}$$

at the boundaries, we get $n \times 2^n$ rectangles such as

| 0 | 0 | 0 |
|---|---|---|
| 0 | 0 | 1 |
| 0 | 1 | 0 |
| 0 | 1 | 1 |
| 1 | 0 | 0 |
| 1 | 0 | 1 |
| 1 | 1 | 0 |
| 1 | 1 | 1 |

Keep in mind that just because an LLL can be described in a local way, large configurations are not necessarily easy to construct. As was thought of quasicrystals up until a local algorithm was discovered [46], there may be no local way to grow large blocks from smaller ones; and attempts to relax large blocks from random initial conditions may lead to very slow, glass-like dynamics, as in some 2- and 3-dimensional models (e.g. [54]). The difficulty of growing a pattern from an initial seed, or relaxing to a pattern from a random initial condition, are themselves good definitions of complexity, and not necessarily correlated with the complexity of recognizing a completed picture.

## 2.3 Deterministic Finite Automata, or DFA's

The automata in the next two sections were introduced by Blum and Hewitt [6]. The reader may consult [51] for a review.

**Definition.** A *4-way deterministic finite-state automaton*, or 4-way DFA, consists of a finite set of states $S$, an initial state $s_0 \in S$, a subset $S_{\text{accept}} \subset S$, and a transition function $F : A \times S \to S \times \{\uparrow, \downarrow, \leftarrow, \rightarrow\}$. $F$ describes how the DFA changes its state and moves one step up, down, left or right, as it encounters symbols in the alphabet $A$. We say that a DFA accepts a block if, starting in the state $s_0$ in the upper-left corner, it eventually reaches some state in $S_{\text{accept}}$; we can demand without loss of generality that this happens at the lower-right corner. We say a 2-d language is *DFA-recognizable* or simply DFA if there exists a 4-way DFA which recognizes it.

As for boundary conditions, we have a choice. The DFA can be *bounded*, in that it must always move back into the block if it detects a $\sharp$, or *unbounded*, in which it is allowed to move into the $\sharp$'s. However, it can be shown that an unbounded DFA can be simulated by a bounded one [39]; if it has $n$ states, it will either return to the block within $n$ steps, or get caught in a loop and wander off to infinity. So these two are equivalent.

We then have

**Proposition.** *The class of DFA languages properly contains the LLL's.*

**Proof.** We first show containment. We simply define a DFA which scans from left to right and top to bottom, and checks around the neighborhood of each site. It checks that each neighborhood is allowed by the LLL, and accepts when it arrives in the lower-right corner; if it finds an illegal neighborhood, it enters a Reject state and 'hangs.'

To show the containment is proper, we give an example which is DFA but not LLL. Consider the language $L_{\text{square}}$ shown in figure 1, of square blocks of 1's on a background of 0's. A DFA can recognize it by scanning the entire scene until it hits the upper-left corner of a block of 1's; it then checks that it is a complete rectangle by scanning its interior, and making sure its edges are straight. Finally, it travels diagonally, up and left, from the lower-right corner; if it arrives at the upper-left corner, it has verified that it is in a square, and moves on to find the next block of 1's.

However, this language is not an LLL. Suppose it were, with range $r$. Then it would be unable to distinguish squares of side greater than $r$ from rectangles, since as shown in figure 1 they contain all the same neighborhoods of size $r$ or less. ∎

As this example shows, a DFA can exploit the geometry of the two-dimensional lattice in order to test for structure that in one dimension would be considered non-regular (context-free or even context-sensitive).

As another example, consider the language of $p \times q$ rectangles of 1's in a sea of 0's, where $p$ and $q$ are mutually prime; a DFA can test for this by bouncing around inside each rectangle like a billiard ball, accepting (and scanning for the next rectangle) if it arrives one site to the left of the corner where it started. Figure 2 shows this and another example, where a DFA makes knights' moves alternately in the directions $(2, -1)$ and $(-1, 2)$ to verify that it is in a square of side $2^n$. These kinds of arithmetic properties would require a context-sensitive grammar to recognize in one dimension.

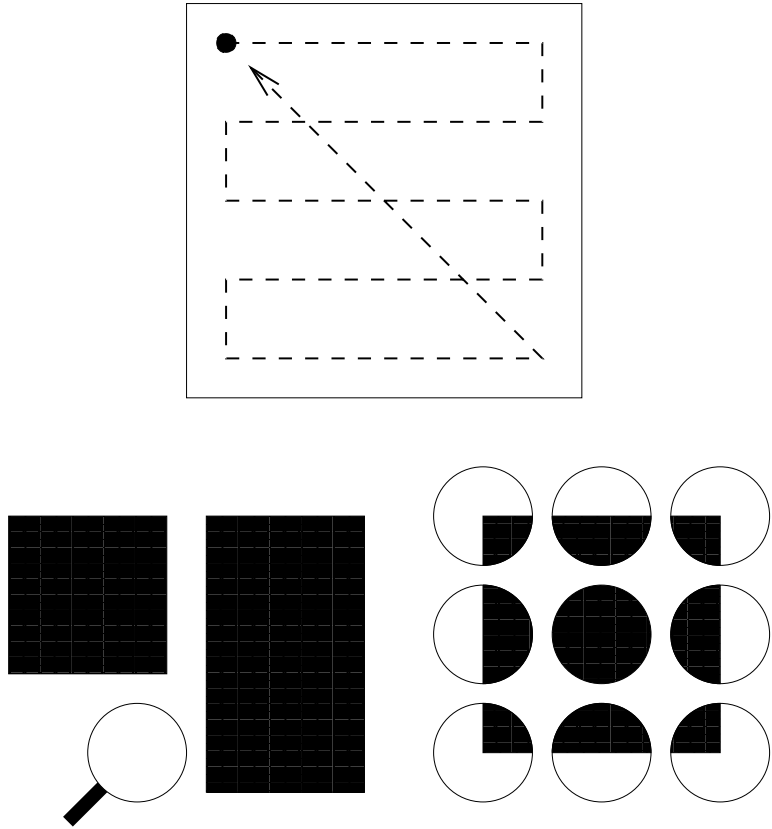

Figure 1: A DFA can recognize the language $L_{\text{square}}$ of square blocks by moving diagonally, but an LLL of range $r$ cannot distinguish between rectangles and squares whose sides are longer than $r$, since all the same neighborhoods of size $r$ or less appear in both.

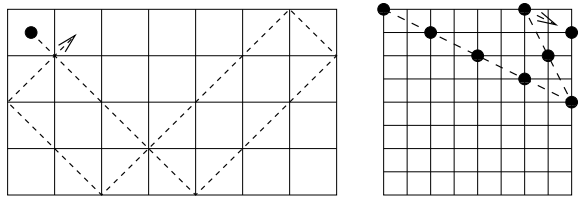

Figure 2: *By bouncing like a billiard ball or making knights' moves, and ending one cell from the corner, a DFA can check that the two sides of a rectangle are mutually prime, or that the side of a square is a power of 2.*

DFAs can get stuck in loops and run forever. However, we can use an argument of Sipser [55] to convert any DFA into one which always arrives in the lower-right corner (in an accepting or non-accepting state), and never gets stuck in a loop. This works by starting in the lower-right corner in an accepting state, doing a depth-first backwards search of the tree of all possible trajectories to see if we could have started in the initial state, and using the DFA's own dynamics to move back up the tree. As a corollary, the complement of a DFA language is also DFA, as we will mention below.

## 2.4 Non-deterministic Finite Automata, or NFA's

The next type of 2-d automaton to consider is the NFA:

**Definition.** A *4-way non-deterministic finite-state automaton*, or 4-way NFA, consists of a finite set of states $S$, an initial state $s_0 \in S$, a subset $S_{\text{accept}} \subset S$, and a non-deterministic transition function $F : A \times S \to \wp(S \times \{\uparrow, \downarrow, \leftarrow, \rightarrow\})$. We say that an NFA accepts a block if there exists a set of choices in $F$ which leads it from the state $s_0$ in the upper-left corner to some state in $S_{\text{accept}}$ (without loss of generality, in the lower-right corner). We say a 2-d language as *NFA-recognizable* or simply NFA if there exists a 4-way NFA which recognizes it.

Recall that in one dimension, DFA's and NFA's are equivalent. In two or more dimensions, NFA's are more powerful:

**Proposition.** *The class of NFA languages properly contains the DFA languages.*

**Proof.** Containment is obvious; we take an example from [51] which is NFA but not DFA. Let the alphabet be $\{0, 1, 2\}$, and consider squares of non-0's on a background of 0's, where the squares have odd side and their center site is a 2.

After confirming that it is in a square as we did before, an NFA can recognize this by moving diagonally from one corner, non-deterministically turning 90° at the 2 in the center, and arriving at another corner.

A DFA, on the other hand, will get lost; there may be many other 2's floating around, and inside a sufficiently large square it has no way of knowing when it is in the center. A counting argument to prove this is given in [51]. ∎

As another example of an NFA language, consider white mazes on a black background, with a red square $a$ and a green square $b$: is there a path from $a$ to $b$? An NFA can non-deterministically guess a path and confirm its existence, but for any DFA (or even DPDA) there is a maze it will get lost in and loop forever [8] (this is an open question if the DFA can move through walls). The "keep your hand on the right-hand wall" method, for instance, will fail if the maze has a loop with $a$ outside and $b$ inside [21]. We will use this as a canonical NFA problem to discuss the computational complexity of NFA languages in Section 2.8 below.

## 2.5 Homomorphisms of LLL's, or h(LLL)'s

We now come to our most subtle class of 2-d languages.

**Definition.** Suppose $L$ is a 2-d language over an alphabet $A$. Then we say $L$ is a *homomorphism of a local lattice language* or h(LLL) if there is some LLL $L'$ in an alphabet $A'$, and a *homomorphism* or mapping $h : A' \to A$, such that $h(L') = L$; i.e., $L'$ yields $L$ when each symbol $a'$ is replaced by its image $h(a')$.

In other words, there is an underlying LLL, some of whose states are hidden by the mapping $h$. For example, consider the LLL whose only allowed $2 \times 2$ blocks are those occuring in

$$\begin{array}{|ccccc|} \hline 0 & 0 & 0 & 0 & 0 \\ 0 & 2 & 1 & 1 & 0 \\ 0 & 1 & 2 & 1 & 0 \\ 0 & 1 & 1 & 2 & 0 \\ 0 & 0 & 0 & 0 & 0 \\ \hline \end{array}, \quad \begin{array}{|cc|} \hline 0 & 0 \\ 0 & 0 \\ \hline \end{array}, \quad \begin{array}{|cc|} \hline 1 & 1 \\ 1 & 1 \\ \hline \end{array}$$

Then the 2's down the diagonal enforce the squareness of each island. If we apply the mapping $h(0) = 0$, $h(2) = h(1) = 1$, we get the language $L_{\text{square}}$ of the previous section; so this is an h(LLL) which is not an LLL.

As another example, consider the "Eight Queens" problem, where queens are placed on a chessboard in such a way that none of them are attacking each other. The reader can easily construct an h(LLL) with symbols $\{Q, 0\}$ and underlying symbols in the power set of $\{\uparrow, \nearrow, \rightarrow, \searrow, \downarrow, \swarrow, \leftarrow, \nwarrow\}$ that get hidden by $h$ and ensure that none of the $Q$'s are on the same row, column, or diagonal. This is not an LLL by the same kind of argument as for $L_{\text{square}}$; since the queens can attack each other over long distances, all the same finite neighborhoods can occur for both attacking and non-attacking configurations.

In a similar vein, it is shown in [10] that homomorphisms of space-time diagrams of 1-d cellular automata are not necessarily given by cellular automata.

So h(LLL)'s are more powerful than LLL's. How much more? To continue the hierarchy, we have

**Proposition.** *The class of h(LLL)'s properly contains the NFA languages.*

**Proof.** We first show containment. The idea is to use the same crossing sequences used in the equivalence of 2-way and 1-way automata in one dimension, recording what sites the NFA visited and in what states; and using the hidden states of the h(LLL) to guess an accepting trajectory for the NFA.

Let $A' = A \times L$, where elements of $L$ are lists of states and directions indicating what state the NFA was in on each visit to that site and which way it moved from there; for instance, $((s_3, \rightarrow), (s_2, \uparrow), (s_6, \leftarrow))$ would indicate that we visited that site 3 times, the first time in state $s_3$ and then moving right, and so on. Since any loop returning to the same site in the same state can be pruned from the NFA's trajectory, we need only consider lists where each state occurs at most once; so $L$ is finite.

Clearly, then, the consistency of the trajectory, its directions and successive states according to the NFA's transition function, can be represented by a LLL on $A'$. Using corner rules, we can require that we start in the upper-left corner in the proper initial state, and end in the lower-right corner in $S_{\text{accept}}$. Finally, let $h$ hide $L$ and project $A'$ onto $A$. So NFA languages are h(LLL)'s.

We now give an example which is h(LLL) but not NFA. Consider the LLL allowing the $2 \times 2$ blocks occuring in

| 0 | 0 | 0 | 0 | 0 | 0 | 0 | 0 |
|---|---|---|---|---|---|---|---|
| 0 | $a$ | $a$ | $a$ | $b$ | $b$ | $b$ | 0 |
| 0 | $A$ | $A$ | $A$ | $B$ | $B$ | $B$ | 0 |
| 0 | 0 | $A$ | $A$ | $B$ | $B$ | 0 | 0 |
| 0 | 0 | 0 | $A$ | $B$ | 0 | 0 | 0 |
| 0 | 0 | 0 | 0 | 0 | 0 | 0 | 0 |

and let $h$ map $A$ and $B$ to 0 while leaving $a$ and $b$ fixed. This leaves us with strings in the 1-d language $\{a^n b^n\}$ on a background of 0's.

Since this language is context-free but not regular [23], it cannot be recognized by a 1-d NFA which moves only in the row containing the string. But a 2-d NFA allowed to move into the 0's above and below that row is no more powerful than a 1-d NFA confined to it, as we will now show.

Write the NFA's trajectory as a series of pairs $(s, d)$ where $s$ is its state and $d \in \{\uparrow, \downarrow, \leftarrow, \rightarrow\}$ its direction. Then the set of trajectories that move above the string's row into the 0's and then return to it form a context-free language, recognized by a PDA that pushes when it sees an $\uparrow$ and pops when it sees a $\downarrow$. Each such trajectory returns to the row with a total horizontal displacement $\Delta x$, equal to the number of $\rightarrow$'s minus the number of $\leftarrow$'s.

Now the *Parikh mapping* maps words to vectors, counting the number of occurences of each symbol: $\pi(w) = (\#_1(w), \#_2(w), \ldots, \#_n(w))$ where $\#_i(w)$ is the number of occurences of the $i$'th symbol in $w$. If $L$ is a context-free language, $\pi(L)$ is a *semilinear set*, i.e. a finite union of sets of the form $\{p+\sum_i^k a_i q_i \,|\, a_i = 0, 1, 2 \ldots\}$ where $p$ and $q_1, \ldots, q_k$ are $n$-dimensional vectors [49].

Since $\Delta x = \#_\rightarrow(w) - \#_\leftarrow(w)$, and since applying a linear functional to a semilinear set yields another semilinear set, the set of possible $\Delta x$'s is a finite union of sets of the form $\{p + nq \,|\, n = 0, 1, 2 \ldots\}$ depending on the initial and final state of the NFA. In other words, it is *eventually periodic*.

Since the set of words whose lengths form an eventually periodic set is a regular language, this means that the NFA's excursions can be simulated by a two-way 1-d NFA confined to the string's row, which travels from the 2-d NFA's starting point to the point of its return, while keeping track of what state it can be in when it gets there. Since a 1-d NFA can only recognize regular languages, and since $\{a^n b^n\}$ is not regular, this h(LLL) is not NFA. ∎

As another example, consider $n \times 2^n$ rectangles of a single symbol; this is an h(LLL) using the counting machine LLL from Section 2.2, but it is shown in [28] that it cannot be recognized by an NFA. In [15] a function is called *recognizable* if the set of $n \times f(n)$ rectangles of a single symbol is an h(LLL); they show that any function that obeys a linear recurrence relation is recognizable.

Our lemma regarding an NFA's excursions off a $n \times 1$ strip appears to be new; it is an open question whether this can be extended to excursions outside an $m \times n$ rectangle [51].

We can now summarize our results by saying

LLL ⊂ DFA ⊂ NFA ⊂ h(LLL)

with each inclusion proper in $d \geq 2$.

There are many interesting examples of h(LLL)'s; here are some of our favorites.

**1.) Two-dimensional L-systems, produced by some expansion rule.** L-systems [33] are languages generated by applying some dilation rule simultaneously everywhere in the string. For instance, the rules $0 \to 01, 1 \to 10$ generate the *Morse sequence* 0110 1001 1001 0110 . . ., and the rules $a \to ab, b \to a$ generate the *Fibonacci sequence abaab aba abaab* . . . that appears in the renormalization theory of circle maps.

In two dimensions, we can expand the lattice by recursively replacing characters with blocks of a certain size, rather like the expansion rules for perfect quasicrystals. The Sierpinski carpet, for instance, is generated by a $3 \times 3$ expansion rule as shown in figure 3.

To construct an h(LLL) for these, we need an underlying hierarchical structure which identifies blocks with their parent blocks on larger and larger scales. Robinson [50, 18] has constructed a set of tiles which does just that, by running H's along the boundaries between blocks and connecting them with the larger H's of their parent block, in such a way as to enforce a hierarchical tree.

By decorating his tiling as shown in figure 4, we can enforce $2 \times 2$ expansion rules such as a two-dimensional version of the Morse sequence; the homomor-phism removes the H's, leaving us just with the symbols on the leaves. We could also make such a rule either locally (with choices made at each site) or globally (with the same choice used everywhere) non-deterministic.

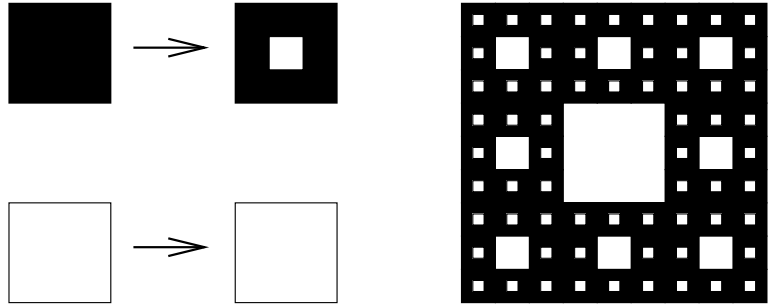

Figure 3: A $3 \times 3$ L-system that generates the Sierpinski carpet.

Clearly, we could make a similar construction for $m \times n$ expansions, as long as $m, n \geq 2$. For a $n \times 1$ rule, which is simply a one-dimensional L-system, we need some blank space above or below the string's row to parse it; for instance, consider the LLL whose allowed $2 \times 2$ blocks are those appearing in

$$\begin{array}{|ccccccccc|}
\hline
\sharp & \sharp & \sharp & \sharp & \sharp & \sharp & \sharp & \sharp & \sharp \\
\sharp & a & a & a & a & a & a & a & \sharp \\
\sharp & a & a & a & a & b & b & b & \sharp \\
\sharp & a & a & b & b & a & a & a & \sharp \\
\sharp & a & b & a & a & a & a & b & \sharp \\
\sharp & A & B & A & B & A & B & A & \sharp \\
\sharp & \sharp & \sharp & \sharp & \sharp & \sharp & \sharp & \sharp & \sharp \\
\hline
\end{array}$$

Then rows of $a$'s have words in $a^*b^*$ below them, and $b$'s have $a$'s below them, enforcing the rules $a \to ab$ and $b \to a$ and ending with Fibonacci sequences of $A$'s and $B$'s at the bottom.

**2.) Topological examples, such as non-simply connected blobs.** Just as an LLL can test local topological properties, an h(LLL) can test for the *existence* of a local topological structure. For instance, consider the LLL on the alphabet $\{0, \nwarrow, \nearrow, \searrow, \swarrow\}$, where we forbid

$$\left\{ \begin{array}{|cc|} \hline \nearrow & \nearrow \\ \swarrow & \swarrow \\ \hline \end{array}, \begin{array}{|cc|} \hline \searrow & \swarrow \\ \nearrow & \nwarrow \\ \hline \end{array}, \begin{array}{|cc|} \hline \nwarrow & \nearrow \\ \swarrow & \searrow \\ \hline \end{array}, \begin{array}{|cc|} \hline 0 & \nearrow \\ \hline \end{array} \right\}$$

and their rotations and reflections. Informally, then, we have a vector field which is outward at the boundary, and has no fixed points except saddle points such as $\begin{array}{|cc|} \hline \searrow & \nearrow \\ \swarrow & \nwarrow \\ \hline \end{array}$ and $\begin{array}{|cc|} \hline \nearrow & \nwarrow \\ \searrow & \swarrow \\ \hline \end{array}$. If $h$ then maps all the arrows onto 1, we get blobs of 1's which support a vector field of this form.

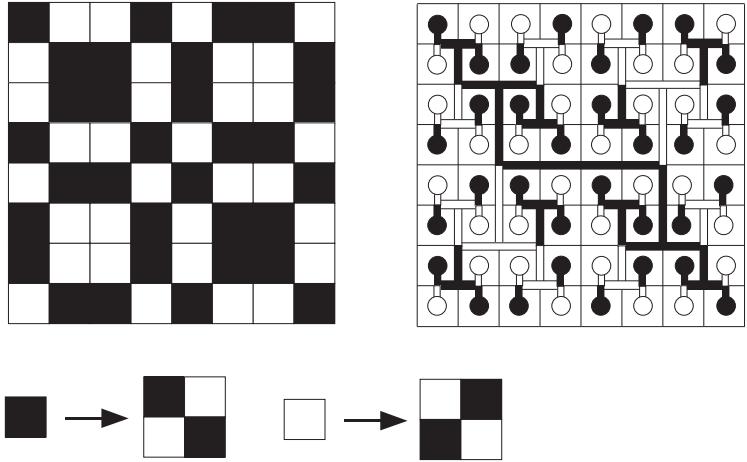

Figure 4: A two-dimensional L-system analogous to the Morse sequence. An h(LLL) can recognize it with a decorated version of Robinson's aperiodic tiles; the parent symbols are carried along the stems of the H's, giving rise to the daughter symbols where they branch.

In the continuous case, the existence of such a vector field on a compact set $U$ in $\mathbb{R}^2$ would imply that $U$ has at least one hole, since by the Poincaré-Hopf index theorem [19] the number of sources, sinks, and circulations minus the number of saddles equals $U$'s Euler characteristic $\chi$. If we allow only saddle points, $\chi \leq 0$, so we have genus 1 or more; if we forbid saddle points as well, $\chi = 0$ and we have exactly one hole.

We now show that this is true in the discrete case as well. Start at the boundary of a finite blob, and trace the vectors backward; since sources are forbidden, there is always at least one predecessor. Since the blob is finite, we must eventually find ourselves on a closed curve.

Suppose the curve is filled entirely with 1's so that there are no holes inside it. Then starting at a point on the curve and heading inward, along either successors or predecessors, we must either come back out or get caught in another cycle, as shown in figure 5. But in either case, we have found a cycle smaller than the first one. We continue this process until we have a cycle around a single vertex, which is forbidden. So by contradiction, the blob contains at least one hole.

Similarly, one can show that two or more holes imply the existence of a 'saddle path', a closed curve connecting four points of the form $a \to b \leftarrow c \to d \leftarrow a$, which contains a smaller saddle path, and so on down to a saddle point. So there can be only one hole per blob if saddles are forbidden.

Conversely, any blob of 1's with one or more holes has a vector field without sources or sinks; just draw a closed curve around each hole, and then extend

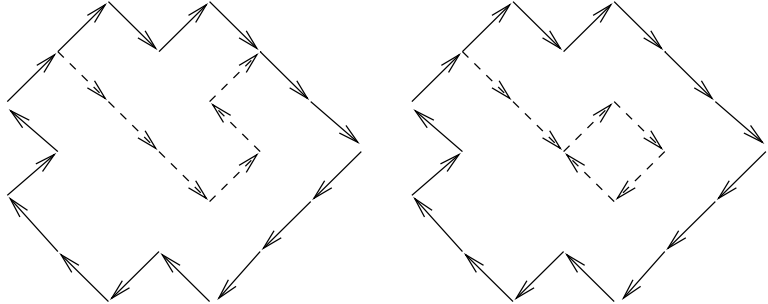

Figure 5: If sources and sinks are forbidden, start at any point in a cycle and move inward (forward or backward along the vector field). The path must re-emerge on the cycle, or get caught in a loop; either way, the cycle contains a smaller one.

arrows outward to the boundaries. (We need a certain thickness to do this; three cells is usually enough.)

Thus the language of blobs of 1's with at least one (or exactly one) hole is an h(LLL). A rather different construction, not based on vector fields, allows us to check whether our blobs of 1's are simply connected, by checking whether the 0's are connected ([51], p. 191).

Nakamura has shown that the language of connected blobs of 1's cannot be recognized by a DFA or NFA in three dimensions [45]; the question is apparently still open in $d = 2$.

**3.) Non-acyclic graphs.** With a suitable alphabet, we can draw out arbitrary directed graphs on a lattice, and then let the underlying LLL guess a cycle in the graph by coloring a set of edges which doesn't begin, end, or branch. Thus the set of directed graphs containing a cycle is an h(LLL).

**4.) NP-complete problems.** Consider an LLL on the alphabet $\{r, g, b, B, 0\}$ in which $r$, $g$, and $b$ represent three colors, 0 a background, and $B$ a boundary between two domains. Then forbid rotations of $\boxed{x \ \ y}$, $\boxed{B \ \ B}$, and $\boxed{x \ \ B \ \ x}$ where $x, y \in \{r, g, b\}$ and $x \neq y$.

If we then apply the homomorphism $h : r, g, b \to x$, we get an h(LLL) of *3-colorable maps*, where blobs of $x$'s can be colored red, blue or green such that the colors are the same within each blob but differ across a boundary of $B$'s. Thus an h(LLL) can determine whether or not a map is 3-colorable.

The reader may be aware that this problem is **NP**-*complete*. Another such problem is Boolean Satisfiability, the question of whether a Boolean circuit has a set of inputs that makes the output true [12]; the reader may enjoy constructing an h(LLL) with symbols for wires and logical gates where the hidden states guess the input values. This is similar to why questions such as whether a spin glass has a ground state below a certain energy are **NP**-complete [2]: the hidden states guess the spin configuration, and $h$ leaves just the couplings visible. We discuss **NP**-completeness, and analogous results for DFA's and NFA's, further in Section 2.8.

Giammarresi and Restivo [13] call h(LLL)'s the *recognizable* languages or REC. They show that they are exactly the homomorphisms of languages definable with a two-dimensional analog of regular expressions, with horizontal and vertical versions of concatenation and the ∗ operator [15]; they also show that h(LLL)'s are exactly the languages definable with existential monadic second-order formulas [14]. Inoue and Nakamura [26] also defined a class equivalent to h(LLL)'s, the *non-deterministic on-line tesselation acceptors*.

## 2.6  Closure properties

One of the most basic questions we can ask about a class $\mathcal{C}$ of languages is whether it is closed under various operations: for instance, for two languages $L_1, L_2 \in \mathcal{C}$, are $L_1 \cap L_2$, $L_1 \cup L_2$, or $\overline{L}_1$ also in $\mathcal{C}$? A class with many such closure properties is a more elegant algebraic object, and more likely to capture a natural set of languages, than one without.

Closure properties can often simplify a proof in automata theory; for instance, if $L = L_1 \cap L_2$, it might be easier to show that $L_1$ and $L_2$ are regular than to show that $L$ is directly.

In one dimension, the regular languages are closed under intersection, union and complement [23]; this generalizes in different ways in $d \geq 2$, as we will now see.

**Proposition.** *The class of LLL's is closed under intersection but not under union or complement.*

**Proof.** For intersection, simply forbid any block which either LLL forbids.

For union, let $L_1$ consist of isolated 1's and $L_2$ of isolated 2's, both on a background of 0's. Then $L_1 \cup L_2$ consists of pictures with either 1's or 2's, but never both; but an LLL of range $r$ cannot distinguish this set of pictures from those with both 1's and 2's separated from each other by $r$ or more. So $L_1 \cup L_2$ is not an LLL.

For complement, let $A = \{0, 1\}$ and let $L$ be the single picture with all 0's. Then $\overline{L}$ includes pictures with isolated 1's; but an LLL that allows 1's separated by an arbitrary width of 0's will also accept L. So $\overline{L}$ is not an LLL. ∎

However, the union of two LLL's with disjoint alphabets $A_1$ and $A_2$ is clearly an LLL: just forbid neighborhoods containing elements of both $A_1$ and $A_2$, so that each picture consists entirely of one or the other.

**Proposition.** *The class of DFA languages is closed under intersection, union and complement.*

**Proof.** For intersection, run the first DFA and then the second, returning to the upper-left corner after the first one accepts. For complement, use the backwards search from the accepting state described above from [55]. Then the union can be written using De Morgan's law, $L_1 \cup L_2 = \overline{\overline{L_1} \cap \overline{L_2}}$. ∎

**Proposition.** *The class of NFA languages is closed under union and intersection.*

**Proof.** For intersection, run one DFA after the other. For union, non-deterministically choose at the outset which one to run. ∎

We use the closure of NFA's under intersection to prove a proposition in Section 3.3 below.

**Proposition.** *The class of h(LLL)'s is closed under union and intersection, but not under complement.*

**Proof.** For intersection, suppose the h(LLL)'s have two underlying LLL's, $L_1$ and $L_2$, and homomorphisms $h_1$ and $h_2$. Then let $L' = L_1 \times L_2$ be the LLL with pairs of states $(s_1, s_2)$ at each site with the two components obeying $L_1$ and $L_2$ respectively, with the additional requirement that $h_1(s_1) = h_2(s_2)$. Then $L_1 \cap L_2 = h'(L')$ where $h'((s_1, s_2)) = h_1(s_1) = h_2(s_2)$.

For union, assume without loss of generality that the alphabets $A_1, A_2$ of the underlying LLL's are disjoint. Then let $L'$ be the union of the two LLL's on the alphabet $A_1 \cup A_2$ where neighborhoods containing elements of both $A_1$ and $A_2$ are forbidden, and let $h'(s) = h_i(s)$ if $s \in A_i$ for $i = 1, 2$.

For complement, consider pictures consisting of a single row of 2's, with rows of 0's and 1's above and below it, such that there is a row above the 2's which is not equal to any of the rows below the 2's. It is easy to see that this is an h(LLL), but it is shown in [56] that its complement is not. ∎

In [13] it is shown that h(LLL)'s are closed under horizontal and vertical versions of concatenation and the ∗ operator; DFA's and NFA's are not [29, 27].

It is an open question whether NFA's are closed under complement. It seems unlikely, since (for instance) the set of mazes with no route from $a$ to $b$ would be NFA. The basic problem is that NFA's are defined with an existential quantifier, ∃ ("there exists") an accepting trajectory; while the complement of such a set is defined with a universal quantifier, ∀. However, we will see below that NFA's could be closed under complement within standard beliefs about complexity classes.

## 2.7 Extensibility of finite blocks

In many cases, we are interested in the set of blocks of an LLL that can be extended to cover the plane; if we are studying a statistical mechanics system on an infinite lattice, for instance, the only finite blocks that are physically relevant are those that appear in infinite configurations.

The set $R(E_\infty(L))$ of finite blocks that are *infinitely extensible*, i.e. that appear in an infinite allowed configuration of a LLL or h(LLL), can be a proper subset of $L$. It can be more or less complex than $L$; for that matter, it can be empty.

For instance, consider the *LLL* where horizontal, vertical and diagonal lines extend across a blank sea, without being allowed to bend, cross or branch:

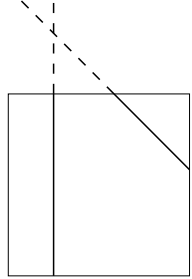

Figure 6: An finite block of the LLL described in the text, which cannot be extended because the two lines intersect.

allowed blocks being of the form 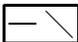, with blocks such as 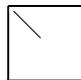 and 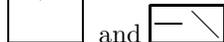 forbidden. A finite allowed block can contain both diagonal and (say) vertical lines; but such a block is not extensible, since those lines intersect outside the block's boundary as in figure 6. The set of extensible blocks consists of those whose lines all have the same orientation; this is recognizable by a DFA.

As another example, consider the h(LLL)'s given above that recognize words in non-regular languages; extensible $n \times 1$ strips of the underlying LLL's are not DFA, NFA or h(LLL) since all of these can only recognize regular one-dimensional languages when confined to a horizontal strip. So the extensible subset of an LLL can have a variety of complexities.

In general, the question of extensibility is undecidable. Recall that a set is *recursively enumerable* if some Turing machine accepts it by halting when given its elements as input, and *recursive* if both it and its complement are recursively enumerable. Then:

**Proposition.** *The set of infinitely extensible finite blocks of an LLL is the complement of a recursively enumerable set, and in two or more dimensions is non-recursive in general. Thus it is undecidable whether a finite block is infinitely extensible.*

**Proof.** To show that its complement is recursively enumerable, consider a Turing machine which takes a finite block as input and attempts to extend it an increasing distance outside its boundary (say, in a spiral around the original block) by doing a depth-first search of possible extensions. If it meets a forbidden neighborhood, it backtracks and tries the next symbol at the most recent place where it had more than one choice; it halts if it has tried all possible states and it has no choices left. So, if the block can only be extended $m$ sites around the spiral, the machine will halt after at most $\mathcal{O}(k^m)$ computation steps where $k$ is the number of symbols in the LLL's alphabet. If the block is extensible, the machine will never halt, so the set of extensible blocks is the complement of the TM's halting set.

To give a non-recursive example, recall that space-time diagrams of one-dimensional cellular automata are LLL's in two dimensions. Choose a CA that simulates a universal Turing machine (e.g. [35]) and forbid any neighborhood containing the halt state. Then the set of extensible $n \times 1$ rows with the Turing machine properly initialized in the upper-left corner are precisely those inputs on which the Turing machine will not halt; this is a non-recursive set since the Halting Problem is undecidable. ∎

The Cluster Variation Method (CVM) in statistical mechanics is a generalization of the mean-field approximation, in which we keep track of the frequency of finite blocks up to a certain size and ignore correlations over larger scales. To apply it, we need to know when there is a measure on infinite configurations that is consistent with a given set of block frequencies. Since only infinitely extensible blocks can contribute to such a measure, it is undecidable in $d \geq 2$ whether the CVM is applicable to a given system [53].

In one dimension, on the other hand, if $L$ is a regular language then its extensible subset is also, since if the finite automaton accepting it has $n$ states, extensibility of a finite word depends only on its first and last $n$ symbols.

The question of whether an LLL in two or more dimensions has any infinite allowed configurations at all is also undecidable; this is the Tiling Problem [5, 50], in which we try to cover the plane with a set of interlocking tiles. In one dimension, a finite state automaton accepts an infinite word if and only if there are loops in its transition graph, which is easily decidable.

Of course, there are classes of LLL's where every block is extensible, such as finite time sets and limit sets of CA's (since by definition these are derived from infinite initial states, see Section 3) and $2 \times 2$ LLL's where the set of allowed blocks is reflection-symmetric, so that a block can be extended as in

$$\begin{array}{|ccc|} \hline a & b & c \\ d & e & f \\ \hline \end{array} \rightarrow \begin{array}{|cccccc|} \hline a & b & c & b & a & \cdots \\ d & e & f & e & d & \\ a & b & c & b & a & \\ \vdots & & & & & \ddots \\ \hline \end{array}$$

In addition, if periodic configurations are dense in the set of infinite allowed configurations, then the extension problem is decidable [57]: since every extensible finite block is contained in a periodic infinite configuration, as we try to extend a block we either run out of choices or reach a periodic block which can be repeated, so either outcome is decided in finite time. This includes the case where the LLL's allowed configurations form a group [31].

## 2.8 Acceptance problems and computational complexity

One way to characterize the power of a class of machines or languages is by the computational complexity of its ACCEPTANCE problem: given a machine $M$ and an $m \times n$ picture $x$, does $M$ accept $x$? Although we have seen that many questions regarding infinite pictures are undecidable, we can get some interesting results if we restrict ourselves to finite ones, and ask how the computational resources needed grow with the size of the picture.

We recommend [48] to the reader as an introduction to the complexity classes we use in the following.

For LLL's, the problem is easy: acceptance is simply the AND of one predicate for each neighborhood, which is true if that neighborhood is allowed. This can be done in parallel in constant time, if we can AND an arbitrary number of things together at once; thus LLL ACCEPTANCE is in the class $\mathbf{SAC}^0$ of problems solvable by *semi-unbounded* constant-depth circuits, where one kind of gate (in this case AND) is allowed to have an arbitrary number of inputs [17].

For NFA's, one might think that a deterministic machine would have to explore an exponential number of trajectories to check for an accepting one. However, NFA ACCEPTANCE is really just a special case of GRAPH REACHABILITY, in which we ask whether there is a path from node $a$ to node $b$ in a directed graph: if the NFA has $s$ states, then the nodes of the graph are the $m \times n \times s$ combinations of location and state, and $a$ and $b$ are the initial and accepting final states in the upper-left and lower-right corners respectively. Conversely, any GRAPH REACHABILITY problem can be converted to an NFA ACCEPTANCE problem by drawing out the graph as a maze, and asking the NFA to find a path from $a$ to $b$ as in Section 2.4 above.

GRAPH REACHABILITY is $\mathbf{NL}$-*complete*, where $\mathbf{NL}$ is the class of problems solvable by non-deterministic Turing machines with logarithmic space; and since these two problems are equivalent, NFA ACCEPTANCE is $\mathbf{NL}$-complete too. Since $\mathbf{NL}$ is contained in the class $\mathbf{NC}^2$ of problems solvable by circuit of depth $\mathcal{O}(\log^2 n)$ for inputs of size $n$, NFA ACCEPTANCE can be solved by a parallel computer in $\mathcal{O}(\log^2 mns)$ time. A serial computer can solve it in $\mathcal{O}(mns)$ time, by starting with the initial state and iteratively adding all possible transitions to a list of accessible states and sites.

In the same way, DFA ACCEPTANCE is a special case of REACHABILITY for directed trees where each node has at most one outgoing edge, and conversely a DFA can explore any such graph drawn on its lattice. This problem is complete for the class $\mathbf{L}$ of deterministic log-space Turing machines, so DFA ACCEPTANCE is $\mathbf{L}$-complete.

Finally, as we saw above, h(LLL) ACCEPTANCE is $\mathbf{NP}$-complete since h(LLL)'s can guess colorings of graphs or satisfying assignments for Boolean expressions; it is in $\mathbf{NP}$ since we can guess the hidden states, and easily check (in $\mathbf{SAC}^0$) whether they satisfy the LLL and match the picture.

Thus we have proved

**Proposition.** *The* ACCEPTANCE *problem is in* **SAC**$^0$, **L**-*complete,* **NL**-*complete, and* **NP**-*complete for LLL's, DFA's, NFA's and h(LLL)'s respectively.*

Since by the Immerman-Szelepcsényi theorem [48] the class **NL** is closed under complement, NFA's could be closed under complement without any drastic consequences to complexity theory.

## 3 Applications to Cellular Automata

### 3.1 Cellular automata

Cellular automata or CA's are spatially extended dynamical systems defined on a regular lattice where the state at each site consists of a symbol from a finite alphabet. Computation theory has been used to characterize the dynamics of one-dimensional CA's (e.g. [44]); in particular one can describe sets of configurations such as periodic sets, finite time sets, and limit sets in terms of their languages of allowed finite blocks. For example, it was shown by Wolfram that any finite time set (set of images allowed at a certain time) of a 1-d CA is described by a regular language [60].

In this Section, we apply some of the results of Section 2 to the dynamics of CA's in two or more dimensions. In particular, we show that the appropriate generalizations of regular languages for fixed and periodic points on the one hand, and finite time sets on the other, are LLL's and h(LLL)'s respectively.

Let us first introduce a few definitions, for simplicity given for the case of a 2-d CA on a square lattice (extensions to higher dimensions and other regular lattices are straight-forward). Let the state at each site $(x, y)$ of $\mathbb{Z} \times \mathbb{Z}$ be $a_{(x,y)} \in A$, where $A$ is a finite alphabet with $|A| = k$ symbols. All lattice sites are updated simultaneously using a local transition function $f : A^B \to A$, where $B$ is some finite neighborhood surrounding each site. This induces a global CA mapping $F : A^{\mathbb{Z} \times \mathbb{Z}} \to A^{\mathbb{Z} \times \mathbb{Z}}$. (According to the Curtis-Hedlund-Lyndon theorem [22], a map on $A^{\mathbb{Z} \times \mathbb{Z}}$ is a CA if and only if it commutes with the horizontal and vertical shift maps, and if it is continuous with respect to product topology induced by the the discrete topology on $A$.)

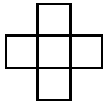

One commonly used neighborhood on the square lattice is the 5-site von Neumann neighborhood shown above; then the local transition function has the form $a_{(x,y)}^{t+1} = f(a_{(x,y)}^t, a_{(x+1,y)}^t, a_{(x-1,y)}^t, a_{(x,y+1)}^t, a_{(x,y-1)}^t)$. A CA has *radius* $r$ if its neighborhood is contained in a square of side $2r + 1$; the von Neumann neighborhood has radius 1.

Typically, we consider the time evolution of *subshifts*, i.e. closed (under the product topology) translation-invariant sets of infinite configurations. Starting with the set $A^{\mathbb{Z} \times \mathbb{Z}}$ of all infinite configurations, only a certain subset of these are allowed after a given number of time-steps; so one basic kind of language we can associate with a CA is the image of $F^t$.

**Definition.** The *finite time set* $\Omega^t$ of a 2-d CA is defined by

$$\Omega^t = F^t(\Omega^0)$$

where $\Omega^0 = A^{\mathbb{Z} \times \mathbb{Z}}$ is the set of all infinite configurations.

As $t \to \infty$, the asymptotic behavior of a CA is described by its limit set, which consists of those configurations in $\Omega^t$ for all $t$ (and conversely, which have predecessors arbitrarily far back in time):

**Definition.** The *limit set* of a CA is defined as

$$\Omega^\infty = \bigcap_{t=0}^{\infty} \Omega^t$$

The finite time sets and limit set clearly obey

$$\Omega^0 \supseteq \Omega^1 \supseteq \Omega^2 \supseteq \ldots \supseteq \Omega^\infty$$

We can also look at the set of periodic points of a CA:

**Definition.** The *period-p set* is given by

$$\Pi^p = \{c \in A^{\mathbb{Z} \times \mathbb{Z}} \,|\, F^p(c) = c\}$$

The set of all periodic points is the *periodic set*, $\Pi = \bigcup_{p=1}^{\infty} \Pi^p$. Clearly $\Pi \subseteq \Omega^\infty$.

These definitions work for infinite configurations. Finite time sets, limit sets, and periodic sets are all subshifts, so equivalent definitions can be given in terms of finite blocks. If we use the local transition function to define a CA map $F : A^{(m+2r) \times (n+2r)} \to A^{m \times n}$ on finite blocks, we have

$$R(\Omega^t) = F^t(R(\Omega^0))$$

and

$$R(\Omega^\infty) = \bigcap_{t=0}^{\infty} R(\Omega^t)$$

where $R$ is the restriction operator defined in Section 2.1.

## 3.2 Periodic sets

Let us first consider the fixed point configurations of a 2-d CA. These form an LLL [47]:

**Proposition.** *The fixed point set of a CA with radius $r$ is described by an LLL of range $2r+1$.*

**Proof.** Simply allow those neighborhoods $\beta \in A^B$ of size $2r+1$ for which the center symbol is fixed, i.e. $f(\beta) = \beta_{(0,0)}$. ∎

Since the $p$'th iteration of the CA mapping is itself a CA with radius $pr$, whose fixed points are the period-$p$ states of the original CA, we also have

**Corollary.** *The period-$p$ set $\Pi^p$ of a CA with radius $r$ is described by an LLL of range $2pr + 1$.*

In fact, the converse is true as well:

**Proposition.** *For any LLL and any $p$, there is a CA for which the LLL is its period-$p$ set $\Pi^p$. Therefore, it is undecidable whether an arbitrary 2-d CA has a periodic orbit of a particular period $p$.*

**Proof.** This is easy in the fixed point case ($p = 1$): let the CA change the value of a site if it belongs to a forbidden block of the LLL, and leave it unchanged otherwise. Then only allowed configurations are fixed.

For $p > 1$, if the LLL has alphabet $A$, introduce an extended alphabet $A' = A \times \{0, \ldots, p\}$. Let the CA rule be given by $(a, n) \to (a, (n+1) \bmod (p+1))$ if the symbol $a$ belongs to a forbidden block, and $(a, n) \to (a, (n+1) \bmod p)$ otherwise. All configurations are then periodic; but one consisting only of allowed blocks has period $p$, while all others have period $p + 1$ or $p(p + 1)$. Thus $\Pi^p$ is the desired LLL.

From the undecidability of the extension problem for LLL's (see Section 2.7), it follows that it is undecidable whether a CA has an orbit of period $p$. ∎

Since homogenous configurations map among each other, a CA with $k$ states always has some periodic orbit of period $p \leq k$. So, in contrast to the result above, the question of whether a CA has <u>any</u> periodic orbits is trivially decidable.

As an example, let us consider the fixed points of the additive 2-d CA defined by
$$a^{t+1}_{(i,j)} = (a^t_{(i,j)} + a^t_{(i+1,j)} + a^t_{(i-1,j)} + a^t_{(i,j+1)} + a^t_{(i,j-1)}) \bmod 2$$
In a fixed point configuration, $a^{t+1}_{(i,j)} = a^t_{(i,j)}$ and
$$a^t_{(i+1,j)} + a^t_{(i+1,j)} + a^t_{(i-1,j)} + a^t_{(i,j+1)} = 0 \bmod 2$$
Then if odd and even sublattices are considered separately, we obtain two independent copies of the LLL in Section 2.2 with an even number of up spins in each $2 \times 2$ block. This means that the entropy per site of the fixed point set is zero, since e.g. for a $n \times n$ diamond (with $n$ sites along diagonal edges and a total of $n^2 + (n-1)^2$ sites) the number of allowed configurations is $N(n) = 2^{4n-4}$.

A more general statement for periodic sets can be made for a class of CA's that form the 2-d analog of the *left (right) permutive* CA rules studied in e.g. [22, 30, 34] (the ergodic properties of this class of 2-d rules are studied in [59]):

**Proposition.** *Consider a 2-d CA with neighborhood $B$ associated with a site $(0, 0)$. Say a site in $B$ is* extremal *if it cannot be written as a convex combination of other points in $B$. If the transition function $f$ is an injective function of some extremal site $(x, y)$ when all other inputs are held fixed, and if $(x, y) \neq (0, 0)$, then the entropy of $\Pi^p$ is zero for any $p$.*

**Proof.** Since $(x, y)$ is extremal, it can be separated from the rest of $B$ by a straight line $L$ as shown in figure 7. Injectivity implies that the value of $a_{(x,y)}$ is uniquely determined by the other sites in the neighborhood together with $a_{(0,0)}^{t+1}$; for a fixed point, $a_{(0,0)}^{t+1} = a_{(0,0)}^t$, so $a_{(x,y)}$ is determined by the rest of the neighborhood. This in turn means that a rectangular block with its top edge oriented along $L$ is uniquely determined by strips of sites of width at most $2r$ along three of its sides; so the number of $m \times n$ configurations grows as $N(m,n) = |A|^{cm+dn}$ for constants $c$ and $d$, and the entropy per site is zero.

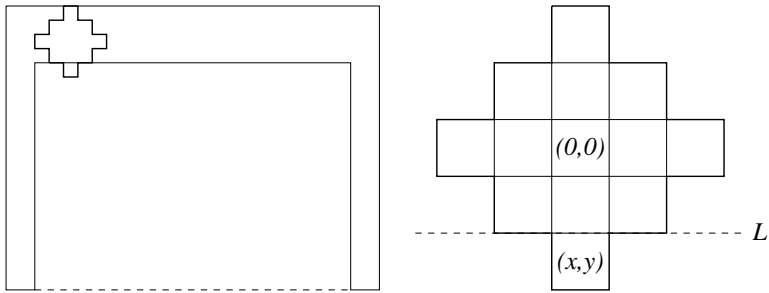

Figure 7: The fixed points of a CA which is injective on an extremal site in its neighborhood are determined by strips of constant width on the outside of a rectangle.

The generalization to $\Pi^p$ follows from the fact that if $f$ is injective on an extremal site $(x, y)$, then $f^p$ is injective on the site $(px, py)$, which is extremal in its larger neighborhood. ∎

This class of CA rules includes additive rules with a prime number of states (except for those with a one-cell neighborhood). Additive CA's with a composite number of states where some periodic sets have positive entropy can easily be constructed, even in $d = 1$: let $k = 4$, and let $a_i^{t+1} = (2a_{i-1}^t + a_i^{t+1}) \bmod 4$. Then any sequence consisting only of symbols 0 and 2 is a fixed point, and all other configurations have period 2; so both $\Pi^1$ and $\Pi^2$ have positive entropy.

## 3.3 Finite time sets

Examples of CA finite time sets can be found in a number of the language classes discussed in Section 2. A very simple example is given by the von Neumann neighborhood rule which maps ┌─┐ to 1 and all other neighborhoods to 0. This rule reaches a fixed point after one time-step, so $\Omega^1 = \Omega^\infty = \Pi^1$ and are all described by the LLL where no adjacent pairs of 1's are allowed.

In general, if the limit set of a CA is described by an LLL, then it is reached at finite time. This was stated in [60] and shown for the 1-d case in [25]. The proof for arbitrary dimensions is essentially identical to that in one dimension, even though this case might seem more subtle because of the distinction between the defining LLL and the set of finite blocks that actually appear in infinite configurations.

**Proposition.** *If the limit set $\Omega^\infty$ of a CA is described by an LLL, there exists a finite time $t$ such that $\Omega^t = \Omega^\infty$.*

**Proof.** If $\Omega^\infty$ is described by a finite list $\Lambda$ of forbidden blocks, then each block $b$ must be excluded at some finite time $t(b)$. If we let $t = \max_{b \in \Lambda} t(b)$, then no further blocks are forbidden after $t$ and $\Omega^t = \Omega^\infty$. ∎

Some LLL's are CA finite time sets (and by this Proposition also limit sets), such as those where some symbol $a$ does not appear in any of the forbidden blocks: let the CA change any symbol to $a$ if it belongs to some forbidden block, and leave it unchanged otherwise. Then after one time-step, only allowed neighborhoods remain.

However, not every LLL is a CA finite time set, as we will now show using a lemma similar in spirit to the Pumping Lemma for 1-d languages:

**The Patching Lemma.** *Suppose $L = \Omega^t$ for some CA with radius $r$. Let $P^1, P^2, \ldots, P^k$ be pictures in $L$. Let $g$ be a function from $\mathbb{Z} \times \mathbb{Z}$ to $\{1, 2, \ldots, k\}$. Then if we define a new picture $P$ in patches as $P_{(x,y)} = P^{g(x,y)}_{(x,y)}$, there is a picture $P'$ in $L$ which coincides with $P$ everywhere where $g$ is constant for $rt$ sites in all directions, i.e. $P$ and $P'$ only differ within $rt$ sites of the boundaries of $g$'s domains.*

**Proof.** (Shorter than the statement of the lemma.) Each $P^i$ is $F^t(Q^i)$ for some initial state $Q_i$. Define $Q$ in patches as $Q_{(x,y)} = Q^{g(x,y)}_{(x,y)}$; then $P' = F^t(Q)$ is as described. ∎

Then we can prove that the language $L_{\text{rect}}$ from Section 2.2 is not a CA finite time set or limit set:

**Proposition.** *There are LLL's which are not CA finite time sets or limit sets.*

**Proof.** Suppose $L_{\text{rect}} = \Omega^t$ for a CA of radius $r$. Let $P_1$ be a single square of 1's of side $n \geq 4rt + 2$, let $P_2$ be all 0's, and let $P$ be a patch of the two as shown in figure 8, where $g = 2$ in a square inside $P_1$ of side $m$ with $n - 2rt > m > 2rt$. Then $P$ has at least some 0's inside a square of 1's, which is clearly not in $L_{\text{rect}}$; so $L_{\text{rect}}$ violates the Patching Lemma.

So $L_{\text{rect}}$ cannot be a finite time set of any CA, and since limit sets which are LLL's are also finite time sets, it cannot be the limit set of any CA either. ∎

On the other hand, we have

**Proposition.** *Any h(LLL) is the intersection of a CA finite time set with an LLL.*

**Proof.** Let the CA's alphabet be the alphabet of the underlying LLL, with an additional "error" symbol $x$. Then let the CA rule map any site belonging to a forbidden block of the LLL to $x$, and map sites whose neighborhoods are allowed according the homomorphism $h$. Then the h(LLL) is the intersection of $\Omega^1$ with the LLL $L_{\overline{x}}$ that forbids $x$ from appearing. ∎

**Corollary.** *There are CA finite time sets that are not NFA's.*

**Proof.** Let $L$ be an h(LLL) which is not NFA, such as the one from Section 2.5 that recognizes strips of $\{a^n b^n\}$. Then $L = \Omega^1 \cap L_{\overline{x}}$ by the previous Propo-sition. The NFA's are closed under intersection (Section 2.6), and $L_{\overline{x}}$ is an LLL and so also an NFA, so $L$ would be an NFA if $\Omega^1$ were, and it isn't. ∎

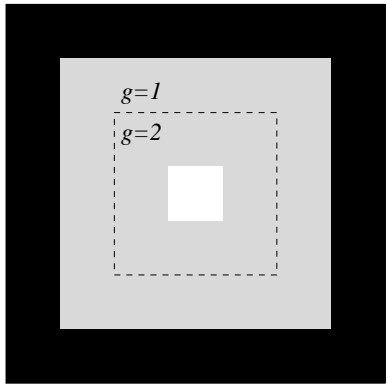

Figure 8: Using the Patching Lemma to show that $L_{\text{rect}}$ is not a CA finite time set. Black and white are 1's and 0's respectively.

To complete the relationship between CA finite time sets and the language classes we have discussed, we have

**Proposition.** *The class of CA finite time sets is properly contained in the class of h(LLL)'s.*

**Proof.** First we show that $\Omega^t$ is an h(LLL) for any $t$. Let the CA's neighborhood be $B$, and denote the neighborhood of the $t$'th iteration of the CA mapping $F^t$ as $B^t$. Introduce a new alphabet $A' = A^{B^t}$, whose symbols consist of neighborhood configurations $\beta$. Then an LLL of range $2rt$ can ensure that the $\beta$'s overlap in a consistent way, and $\Omega^t$ is obtained by applying the homomorphism $h = F^t$.

To show the inclusion is proper, take $L_{\text{rect}}$ or any other h(LLL) that is not a CA finite time set. ∎

One important property of CA finite time sets (and also CA limit sets) is that they are, by definition, extendable to infinite configurations since they are generated from infinite initial conditions; thus the extension problem is trivially decidable. In addition, we can show that the growth function $N(m,n)$ of a CA finite time set must have a comparatively simple form.

In Section 2.2 we saw LLL's in $d=2$ with a wide variety of $N$, including a number with zero entropy. This is in sharp distinction to the one-dimensional case, in which the leading behavior of $N(n)$ for regular languages is always $n^k \lambda^n$ where $\lambda$ is algebraic and $k$ is a non-negative integer, and the entropy is $\sigma = \log \lambda$. For CA finite time sets we can recover a weak analogy of this:

**Proposition.** *In any number of dimensions, a finite time set $\Omega^t$ of a CA with radius $r$ either consists of one homogeneous picture, or has a growth func-* *tion for a volume v bounded by*

$$N(v) \geq 2^{v/(2rt+1)^d}$$

*and thus an entropy per site of at least $\sigma \geq (\log 2)/(2rt+1)^d$.*

**Proof.** Use the Patching Lemma; unless $L$ contains only one picture, two neighborhoods get mapped to different states by $F^t$, and we can choose to fill each block of width $2rt+1$ with one or the other, giving the stated result. ∎

Thus the entropy of $\Omega^t$ cannot decrease faster than $t^{-d}$ unless the CA converges to a single homogeneous fixed point in finite time.

## 3.4  Limit sets

In [24], Hurd uses travelling particles to enforce context-free and context-sensitive structures in the limit sets of one-dimensional CA's. We can use a similar strategy to construct limit sets in two dimensions which are DFA, NFA, or h(LLL).

The CA rule sketched in figure 9, for instance, is designed to allow squares of 1's in a sea of 0's, by extending a string of $a$'s down and to the right from the upper-left corner of each rectangle of 1's. If it meets another corner, it knows the rectangle is a square; it turns the head of the string to a $b$, retracts it up and to the left, and begins again. If it meets an edge it generates an error symbol $x$ (which is also generated if the rules of $L_{\text{rect}}$ are violated), which propagates at the speed of light and destroys the entire lattice.

This CA's limit set then consists of squares of 1's, with strings of $a$'s and an optional $b$ at various stages of construction along the diagonal, plus various propagating fronts of $x$'s. It is easy to show that this is a DFA but not an LLL, since we get $L_{\text{square}}$ if we intersect it with the LLL forbidding $a$, $b$, and $x$.

Similarly, we can recognize squares of 1's and 2's with 2's in the center by extending synchronized strings along two diagonals, or strips of the language $\{a^n b^n\}$ by extending strings from the middle and the ends as shown in figure 10. Then the corresponding limit sets are NFA and h(LLL) respectively.

Every CA limit set is the complement of a recursively enumerable set, but is not generally recursive, even in one dimension [24]. The proof is similar to that for extensibility in Section 2.7, except that the Turing machine tries to construct preimages, rather than extensions, of blocks to see if they are in $R(\Omega^\infty)$.

We used the Patching Lemma above to prove that there are LLL's which are not limit sets. A logical question, then, is

**Open question:** Are there DFA's, NFA's or h(LLL)'s, other than LLL's, which we can prove are not CA limit sets?

# 4  Generalizations

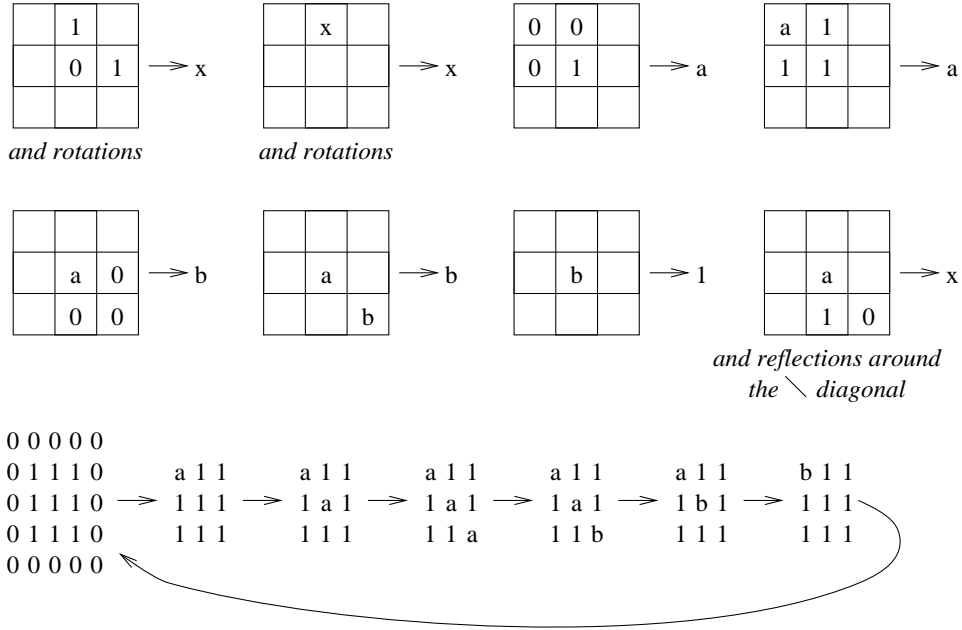

Figure 9: A CA rule on the alphabet $\{0, 1, a, b, x\}$ that allows only squares of 1's on a background of 0's in its limit set, and its evolution on a $3 \times 3$ square. A blank means "don't care", and the $x$-spreading rule takes precedence over all others.

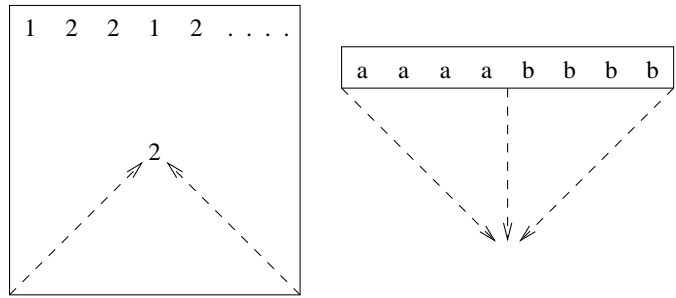

Figure 10: Enforcing NFA and h(LLL) limit sets with synchronized particles.

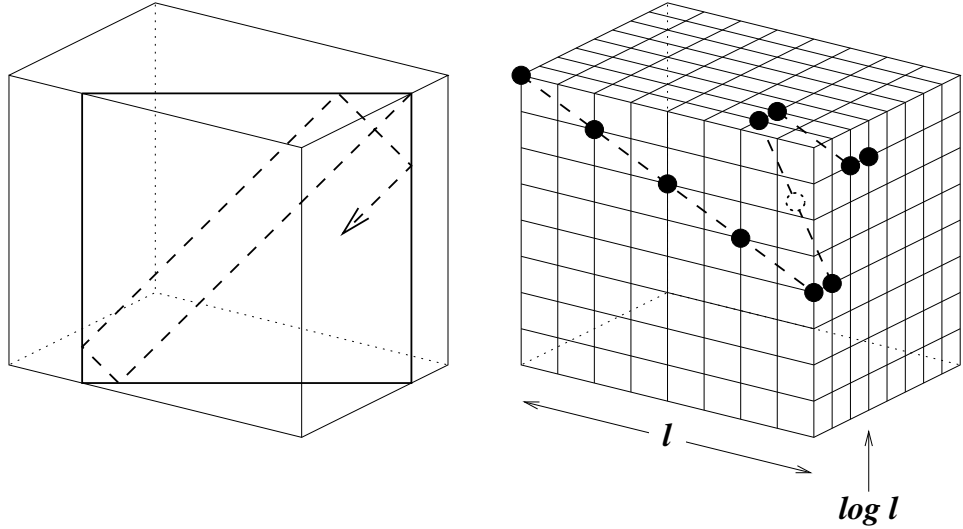

Figure 11: By using the method of figure 2 on diagonal slices, a 3-d DFA in a cube of side $l$ can confirm that $m$ and $l$ are mutually prime for all $m < l$, so that $l$ is prime. By generalizing the knights' move method, it can calculate $\log l$, and then iterate this process to confirm that $l = 2 \uparrow_3 k$ for some $k$.

## 4.1 Higher dimensions

Going to three or more dimensions doesn't change things as much as going from one to two did. The proper inclusions LLL $\subset$ DFA $\subset$ NFA $\subset$ h(LLL) still hold: $d$-dimensional cubes are DFA but not LLL, odd-sided cubes of 1's and 2's with a 2 in the center are NFA but not DFA, and h(LLL)'s that are strips or layers of lower-dimensional non-NFA languages surrounded by blanks cannot be recognized by NFA's. The results on CA languages in Section 3 also hold.

However, there are some interesting higher-dimensional examples. Cubes of prime side $l$ are 3-d DFA; by taking diagonal slices as shown in figure 11, we can use the 2-d DFA of figure 2 to confirm that $l$ and $m$ are mutually prime for every $m < l$. If we have a fourth coordinate, we can add $m$ to it whenever $m$ divides $l$, and check that $l$ is the sum of its factors; so tesseracts of perfect side are 4-d DFA!

We can also generalize the $2^n \times 2^n$ square language discussed above. If we define
$$2 \uparrow_n k = \underbrace{2 \uparrow_{n-1} (2 \uparrow_{n-1} (\ldots (2 \uparrow_{n-1} 2)))}_{k \text{ times}}, \quad 2 \uparrow_0 k = 2 + k$$
then $2 \uparrow_1 k = 2k$, $2 \uparrow_2 k = 2^k$, and so on. Then a DFA in $d \geq 1$ dimensions can verify that it is in a $d$-cube of side $l = 2 \uparrow_d k$ for some $k$. In $d = 3$, for example, it does this as shown in figure 11 by moving one step in a perpendicular direction at the end of each series of knights' moves; at the end of the process this coordinate is $\log_2 l$. We can then use this as a starting point and calculate $\log_2 \log_2 l$, and so on, until we get to 1, showing that $l = 2 \uparrow_3 k = 2^{2^{\cdot^{\cdot^{2}}}}$ for some number $k$ of levels of exponentiation. If we define llog $l = k$, then in $d = 4$ we can calculate iterated llogarithms, and so on.

For fun, if we define

$$3!_n k = \underbrace{3!_{n-1}(3!_{n-1}(\ldots(3!_{n-1}1)))}_{k \text{ times}}, \ 3!_1 k = 3 \underbrace{!!! \ldots !}_{k \text{ times}}$$

then we state without proof (exercise for the reader) that a DFA in $d \geq 4$ can check that a cube has side $3!_{d-3} k$ for some $k$.

The DFA's in this and the previous example can be thought of as finite automata with $d$ counters, namely the DFA's coordinates; it increments and decrements them by moving, and checks for zero by hitting the side of the cube.

Topological examples also become more interesting. By checking for the existence of a consistent normal, 3-d h(LLL)'s can confirm that a manifold is orientable; and perhaps a clever reader can come up with a discrete foliation or vector field in $\mathbb{R}^3 - K$ which only exists for certain knot or link types $K$. We can also more easily represent non-planar graphs.

However, some problems may get harder: the "keep your hand on the left-hand wall" algorithm for traversing acyclic mazes no longer works for $d > 2$, and some spin systems that are exactly solvable in two dimensions are not known to be in three.

## 4.2 Higher types of acceptors

Why not continue up the ladder of the Chomsky hierarchy, to two-dimensional versions of push-down automata and Turing machines? Partly because the distinction between one and more dimensions is not so great as for regular languages. Recall that a *two-way push-down automaton* (2PDA) is a finite-state machine with access to a stack memory which can move left or right on its input; a *Turing machine* (TM) is a finite-state machine which can move left or right and write new symbols on its tape; and a *bounded Turing machine* is one which is confined to the part of the tape its input is written on. Bounded TM's accept the *context-sensitive* languages [23]. Then:

**Definition.** If $w$ is an $m \times n$ picture, let raster($w$) be $w$'s rows separated by marker symbols, $w_{(1,1)} \ldots w_{(1,n)} \natural w_{(2,1)} \ldots w_{(2,n)} \natural \ldots \natural w_{(m,1)} \ldots w_{(m,n)}$. If $w$ is a $d$-dimensional picture, we separate its rows with $d-1$ different marker symbols $\natural_1, \ldots \natural_{d-1}$.

**Proposition.** *If $L$ is a $d$-dimensional language recognizable by a $d$-dimensional (PDA, TM, bounded TM), then raster($L$) is recognizable by a one-dimensional (2PDA, TM, bounded TM).*

**Proof.** We will prove this for $d = 2$; the generalization to higher dimensions is straightforward.

We need to show that the one-dimensional version of each of these machines can simulate its two-dimensional version, by moving "up" (resp. "down") in raster($w$) from $w_{i,j}$ to $w_{i-1,j}$ (resp. $w_{i+1,j}$). A PDA can do this as follows: scan for the first ♮ to the left of your current position, pushing a symbol $x$ onto the stack at each step; there are now $j$ $x$'s on the stack. Then move to the next ♮ to your left (right), and then move right, popping an $x$ at each step, until there are no $x$'s left. You are now in the $j$'th site in the row to the left (right) of your original position.

A Turing machine can accomplish the same thing by marking its current position with an $a$ and marking the next ♮ beyond the one to its left (resp. the ♮ to its right) with a $b$, and then shuttling back and forth, moving the $a$ to the left and the $b$ to the right, until after $j$ steps the $a$ arrives at the ♮ to the left of its original position. The $b$ is now $j$ sites to the right of the appropriate ♮. ∎

Surprisingly, ACCEPTANCE for deterministic two-way PDA's in one dimension is decidable in linear time [7], so we have

**Corollary.** ACCEPTANCE *for deterministic PDA's in any number of dimensions can be decided in time proportional to the volume.*

There are several higher types of recognizers that the reader should be aware of if she wishes to further explore this subject, such as:

**Alternating Finite Automata (AFA's).** These are a generalization of NFA's in which the tree of possible trajectories can have both *existential* nodes, that require at least one of their subtrees to accept, and *universal* nodes, that require all of their subtrees to accept (NFA's are just AFA's with only existential nodes). In one dimension AFA's only recognize regular languages, but in $d \geq 2$ they are more powerful than NFA's; the relationship between h(LLL)'s and AFA's is an open question [29]. The ACCEPTANCE problem for AFA's is **P**-complete [17], suggesting that they lie between NFA's and h(LLL)'s in power.

**Pebbling automata.** These are finite-state automata that have a fixed supply of *pebbles*, which they can pick up or deposit on sites of the input, and sense when they run across them. In one dimension, one-pebble machines can only recognize regular languages, even in the alternating case [6, 16]; in two dimensions, the reader can easily show that both the NFA language of squares with a 2 in the center and the h(LLL) of strips of $\{a^n b^n\}$ can be recognized by a one-pebble DFA, and an NFA with one pebble can look for cycles in a graph.

**Multi-head finite automata.** These are finite-state automata with multiple heads which they can move independently on the input. A two-head DFA, for instance, can recognize the language $\{w\natural w\}$ of words repeated twice with a marker in the middle, which is neither regular nor context-free. If DFA(k) and NFA(k) are the classes recognized by $k$-head DFA's and NFA's, both of these form distinct hierarchies (i.e. $k + 1$ heads are more powerful than $k$) in $d = 1$ and therefore in higher dimensions as well [42]. A logical characterization of $k$-head DFA and NFA languages is given in [3].

We could also discuss *counter h(LLL)'s* where the hidden states contain one or more integers, with an underlying local rule that can impose inequalities between them and those of their neighbors, increment or decrement them from neighbor to neighbor, or check for zero. Since these can recognize non-regular languages in one dimension such as the Dyck language, they properly contain the h(LLL)'s. A counter h(LLL) could also check that a vector field is a gradient, or that a directed graph is acyclic, by assigning an altitude $s$ at every site such that $s_i > s_j$ if there is an edge $i \to j$. We believe, but have not been able to prove, that ordinary h(LLL)'s cannot recognize this language; they can recognize its complement, however, as we showed in Section 2.5.

## 5 Conclusion

We have shown that the notion of "regular language" generalizes in several different ways in two or more dimensions: LLL's, DFA's, NFA's and h(LLL)'s. The examples given hopefully give the reader an intuition for what each class is capable of.

As tools and applications, we have studied the closure properties of these classes, related their ACCEPTANCE problems to the complexity classes $\mathbf{SAC}^0$, $\mathbf{L}$, $\mathbf{NL}$ and $\mathbf{NP}$, and applied them to the languages generated by cellular automata in finite and infinite time.

We hope that we have given the reader some conceptual tools for the classification of two-or-more-dimensional patterns found in her research; or at least that she has found the examples and distinctions we have made enjoyable.

A preliminary version of this work appeared in [36].

**Acknowledgements.** We are grateful to the Niels Bohr Institute where this work was begun, the Santa Fe Institute where it was continued, and the Bellairs Research Institute of McGill University where it was finally put to rest. We also thank Jean-Camille Birget, Marek Chrobak, Tao Jiang, Oliver Matz, and David Simplot for helpful communications. This work was supported in part by NSF grant ASC-9503162.

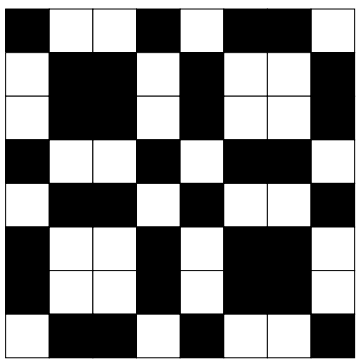
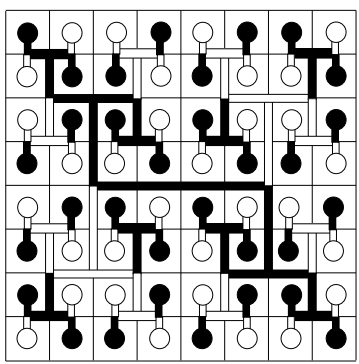
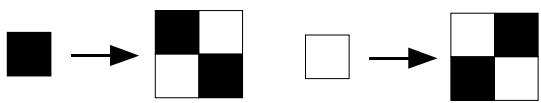